\documentclass[12pt]{article}
\usepackage{amsmath,amssymb}
\usepackage{graphics}
\usepackage[utf8]{inputenc}
\usepackage[T1]{fontenc}
\usepackage[colorlinks=true, linkcolor= PersianBlue, citecolor= PersianBlue, urlcolor=CarmineRed, bookmarks=true, bookmarksopen=true]{hyperref}
\usepackage[a4paper, left=2.75cm, right=2.75cm, top=2.75cm, bottom=3.5cm]{geometry}
\usepackage{color}
\usepackage{braket}
\usepackage{txfonts}
\usepackage{verbatim}
\definecolor{CarmineRed}{rgb}{1,0,0.2}
\definecolor{PersianBlue}{rgb}{0.1098,0.22353,0.73333}
\newcommand{\beq}{\begin{equation}}
\newcommand{\eeq}{\end{equation}}
\newcommand{\diff}{\mathrm{d}}
\newcommand{\Tr}{\mathrm{Tr}}
\newcommand{\GeV}{\mathrm{\ GeV}}
\def\bea{\begin{eqnarray}}
\def\eea{\end{eqnarray}}
\usepackage{feynmp}
\usepackage{subfigure}
\usepackage{epsfig}
\usepackage{setspace}
\usepackage[margin=0pt,font=small,labelfont=bf,textfont=it]{caption}
\usepackage[toc,page]{appendix}

\begin{document}\setlength{\unitlength}{1mm}
\begin{titlepage}
\flushright
TUM-HEP 762/10\\
MPP-2010-68

\vspace{2truecm}
\begin{center}
\boldmath

{\Large\textbf{Radiatively induced flavour violation in the general two-Higgs doublet model with Yukawa alignment}}\\

\unboldmath
\end{center}

\vspace{0.3truecm}
		
\begin{center}
{Carolin B.~Braeuninger$^{\bigstar, \varheartsuit}$, Alejandro Ibarra$^{\bigstar}$ and Cristoforo Simonetto$^{\bigstar}$}
\vspace{0.4truecm}

{\footnotesize
$^\bigstar${\sl Physik-Department T30d, Technische Universität München,\\ James-Franck-Straße, 85748 Garching, Germany}\vspace{0.2cm}

$^\varheartsuit${\sl Max-Planck-Institut für Physik (Werner-Heisenberg-Institut),\\ Föhringer Ring 6,
80805 München, Germany}\vspace{0.2cm}}

\vspace{0.3truecm}

June 14, 2010

\end{center}
\vspace{0.6cm}

\date{May 31, 2010}

\onehalfspacing
\begin{abstract}
The most general two Higgs doublet model contains new sources of flavour violation that are usually in conflict with the experimental constraints. One possibility to suppress the exotic contribution to the flavour changing neutral currents consists on imposing the alignment of the Yukawa couplings. This condition presumably holds at a high-energy scale and is spoiled by the radiative corrections. We compute in this letter the size of the radiatively induced flavour violating Higgs couplings at the electroweak scale. These also yield the absolute lower bound on the size of the exotic contributions to the flavour changing neutral currents in any two Higgs doublet model, barring cancellations and the existence of discrete symmetries. We show that these contributions are well below the experimental bounds in large regions of the parameter space.
\end{abstract}
\end{titlepage}
\onehalfspacing
\section{Introduction}

The LHC finally set out to find the Higgs and in anticipation of its results it is important to consider all viable scenarios for the Higgs sector in order to be able to interpret the data once they arrive. The simplest extensions of the SM Higgs sector are so-called two Higgs doublet models (2HDM) where a second Higgs doublet is added to the SM particle content. In addition to the three Goldstone bosons needed to generate the gauge boson masses,  there are then five physical Higgs particles: two neutral scalars $h$ and $H$, one neutral pseudo-scalar $A$ and two charged scalars $H^\pm$.

In the most general 2HDM there are two Yukawa matrices per fermion type (up-type quarks, down-type quarks, leptons) which cannot be simultaneously diagonalized. Without any further protection, this leads to unacceptably large flavour changing neutral currents (FCNCs).
Therefore in the literature mostly 2HDMs with a discrete symmetry have been considered, ensuring that each right-handed fermion field can couple to no more than one Higgs doublet. There are several choices of the discrete symmetry. The ones most often studied are referred to as type I (only one of the Higgs doublets couples to the fermions) or type II (down-type quarks and leptons couple to one Higgs doublet, up-type quarks to the other) 2HDM. FCNCs are completely absent at tree level in such models.

A generic way to suppress the FCNCs consists in imposing the hypothesis of Minimal Flavour Violation to the flavour symmetry breaking parameters \cite{D'Ambrosio:2002ex}. This hypothesis is implemented in the aligned two-Higgs Doublet model recently discussed in \cite{Pich:2009sp}: If the two Yukawa couplings for each fermion type are aligned, they are simultaneously diagonalizable and FCNCs are absent at tree level. This ansatz is more general than discrete symmetries in two ways: Firstly, it contains type I and type II and all other 2HDMs with discrete symmetries as particular cases. Secondly, it allows to have additional sources of CP violation in the Yukawa sector, in contrast to type I and II 2HDM.

The alignment condition presumably holds at a high-energy scale and, in general, will be spoiled by quantum corrections. In this letter we analyse the size of these corrections in order to determine the viability of this scenario. Furthermore, in the absence of cancellations and in the absence of {\it ad hoc} discrete symmetries, this scenario yields the lower bound on the exotic contributions to the FCNCs in any 2HDM.

The paper is organized as follows: In sec. \ref{2HDMwithYukAl} we recapitulate the 2HDM and Yukawa alignment. In sec. \ref{RGEs} we present an approximate solution to the renormalization group equations (RGEs) in a 2HDM with Yukawa alignment. Therefrom we derive expressions for
the flavour violating neutral Higgs couplings in the quark sector in sec. \ref{FVcouplings}. Bounds on the parameters entering in these couplings are then derived from experimental constraints on meson-antimeson mixing and the leptonic B decay in sec. \ref{Experiments}. Finally we conclude in sec. \ref{Conclusions}.

\section{The 2HDM with Yukawa alignment}
\label{2HDMwithYukAl}

The general 2HDM we consider has the same fermion content and ${\rm SU}(3)_C \times {\rm SU}(2)_L \times {\rm U}(1)_Y$ gauge symmetry as the Standard Model. The Higgs sector consists of \textit{two} scalar ${\rm SU}(2)$ doublets $\phi_1$ and $\phi_2$ with weak hypercharge $Y=\frac{1}{2}$. Both Higgs doublets couple to all fermions\footnote{This is sometimes called a type III 2HDM.} and the Yukawa part of the Lagrangian therefore reads:
\beq
\begin{split}
\mathcal{L}_{\rm Yukawa}=(Y^{(1)}_u)_{ij}\bar{q}_{Li}'u_{Rj}'\tilde{\phi}_1+(Y^{(1)}_d)_{ij}\bar{q}_{Li}'d_{Rj}'\phi_1+(Y^{(1)}_e)_{ij}\bar{l}_{Li}'e_{Rj}'\phi_1\\
+(Y^{(2)}_u)_{ij}\bar{q}_{Li}'u_{Rj}'\tilde{\phi}_2+(Y^{(2)}_d)_{ij}\bar{q}_{Li}'d_{Rj}'\phi_2+(Y^{(2)}_e)_{ij}\bar{l}_{Li}'e_{Rj}'\phi_2+\rm{h.c.}
\label{YukawaLagrangian}
\end{split}
\eeq
where $i,j$ are flavour indices and $\tilde{\phi}_a=i\tau_2\phi^*_a$. The neutral components of the two Higgs doublets acquire vacuum expectation values (vevs) during electroweak symmetry breaking (EWSB) which in general can be complex. 
While one phase can be rotated away, the phase difference is physical. Nevertheless we can choose to work in a basis where both vevs are positive, shifting the phase to the potential and the Yukawas:
\beq
\left\langle\phi_a\right\rangle=\frac{1}{\sqrt{2}}\left(\begin{array}{c}0\\v_a\end{array}\right).
\eeq
In order to minimize the FCNC effects without imposing {\it ad hoc} discrete symmetries we postulate that at a high energy cut-off scale, $\Lambda$, the Yukawa couplings of the same fermion type are aligned~\cite{Pich:2009sp}. We parameterize this condition as:
\begin{eqnarray}
Y_{u}^{(1)}(\Lambda)&=\cos\psi_uY_{u}, \ \ \  \ \ \ Y_{u}^{(2)}(\Lambda)&=\sin\psi_uY_u,\label{ualign}\\
Y_d^{(1)}(\Lambda)&=\cos\psi_dY_d, \ \ \  \ \ \ Y_d^{(2)}(\Lambda)&=\sin\psi_dY_d,\\
Y_e^{(1)}(\Lambda)&=\cos\psi_eY_e, \ \ \  \ \ \ Y_e^{(2)}(\Lambda)&=\sin\psi_eY_e;\label{ealign}
\end{eqnarray}
The type I 2HDM is contained in this parameterization as the special case $\psi_u=\psi_d=\psi_e=0$ and type II as $\psi_u=0,\ \psi_d=\psi_e=\frac{\pi}{2}$.

\section{Radiative corrections to the aligned Yukawa couplings}
\label{RGEs}

The renormalization group equations (RGEs) for a general 2HDM with Yukawa alignment have been derived in \cite{Ferreira:2010xe}. We reproduce them in our notation in appendix \ref{RGEformulae}. The radiative corrections introduce a misalignment of the Yukawa couplings at low energy. To see whether this leads to unacceptably large FCNCs, we solved the RGEs numerically and analytically using the so-called "leading log approximation" which estimates the down-type quark couplings at the electroweak scale as:
\begin{equation}
Y_d^{(k)}(m_Z)\approx Y_d^{(k)}(\Lambda)+\frac{1}{16\pi^2}\beta_{Y_d^{(k)}}(\Lambda)\log\left(\frac{m_Z}{\Lambda}\right),
\end{equation}
and similarly for the Yukawa matrices of the up-type quarks and leptons. Inserting the $\beta$-function \eqref{betadown}, the coupling at the EW scale takes the form:
\beq
Y_d^{(k)}(m_Z)\approx k_d^{(k)}Y_d+\epsilon_d^{(k)}Y_uY_u^{\dagger}Y_d+\delta_d^{(k)}Y_dY_d^{\dagger}Y_d,
\label{EWcoupling}
\eeq
where the coefficients $k_d^{(k)}$, $\epsilon_d^{(k)}$ and $\delta_d^{(k)}$ can be found in appendix \ref{lowscaleYukawa}, as well as the corresponding formulae for up-type quarks and leptons.

\section{Flavour violating neutral Higgs couplings}
\label{FVcouplings}

To derive the low energy Lagrangian it is convenient to rotate the Higgs fields to a basis where only one of the two doublets, say $\Phi_1$, gets a vev:
\beq
\left(\begin{array}{c}
\Phi_1\\
-\Phi_2
\end{array}\right)
=
\left(\begin{array}{cc}
\cos\beta&\sin\beta\\
\sin\beta&-\cos\beta
\end{array}\right)
\left(\begin{array}{c}
\phi_1\\
\phi_2
\end{array}\right).
\eeq
Here $\tan\beta = \frac{v_2}{v_1}$ is the ratio of the vacuum expectation values in the original basis. In the new basis the Lagrangian can be written in the following form (quark sector):
\beq
\mathcal{L}_{\rm Yukawa}=\frac{\sqrt{2}}{v}\left\{\bar{q}_L'\left(M_u\tilde{\Phi}_1+\Gamma_u\tilde{\Phi}_2\right)u'_R+\bar{q}_L'\left(M_d\Phi_1+\Gamma_d\Phi_2\right)d'_R+\rm{h.c.}\right\}
\label{Lagragian-low},
\eeq
where $v^2=v_1^2+v_2^2=(246 \GeV)^2$ and the couplings $M_{u,d}$ and $\Gamma_{u,d}$ are evaluated at the scale $m_Z$. Their expression in terms of the original couplings in the basis $\{\phi_1,\phi_2\}$ is:
\begin{align}
M_{d,u}(m_Z)&=\frac{v}{\sqrt{2}}\left(\cos\beta\ Y_{d,u}^{(1)}(m_Z)+\sin\beta\ Y_{d,u}^{(2)}(m_Z)\right)\label{M-at-MZ},\\
\Gamma_{d,u}(m_Z)&=\frac{v}{\sqrt{2}}\left(-\sin\beta\ Y_{d,u}^{(1)}(m_Z)+\cos\beta\ Y_{d,u}^{(2)}(m_Z)\right)\label{Gamma-at-MZ}.
\end{align}
In order to rewrite the Lagrangian eq.~\eqref{Lagragian-low} in terms of the mass eigenstates, we first express the Higgs doublets $\Phi_1$, $\Phi_2$ in terms of the physical Higgs fields $h, H, A, H^\pm$ and the Goldstone bosons $G^0,G^\pm$:
\beq
\Phi_1=
\left(\begin{array}{c}
G^+\\
\frac{1}{\sqrt{2}}(v+\cos(\alpha-\beta)H-\sin(\alpha-\beta)h+iG^0)
\end{array}\right),
\eeq
\beq
\Phi_2=
\left(\begin{array}{c}
H^+\\
\frac{1}{\sqrt{2}}(\sin(\alpha-\beta)H+\cos(\alpha-\beta)h+iA)
\end{array}\right),
\eeq
for a CP conserving Higgs potential and where $\alpha$ is the mixing angle of the mass eigenstates. Note that in a general 2HDM the ratio of the expectation values $\tan\beta$ has no well defined meaning. The basis of the Higgs fields can be freely chosen and we could just have started in the basis $\Phi_1,\Phi_2$ instead of $\phi_1,\phi_2$ (thus setting $\beta=0$). The only relevant mixing angle is therefore $\alpha-\beta$. This is in contrast to 2HDM type I and II where there is a clear distinction of the two Higgs doublets by the way they couple to the fermions. The ratio of the two vevs then gets a real, physical meaning.

Finally, we perform unitary transformations in flavour space of the quark fields
\begin{align}
u_L'&=V_u^L(m_Z)\ u_L, & u_R'&=V_u^R(m_Z)\  u_R,\\
d_L'&=V_d^L(m_Z)\  d_L, & d_R'&=V_d^R(m_Z)\  d_R, \label{dVd}
\end{align}
in order to diagonalize the quark mass matrices:
$M_u=V_u^{L} M_{u}^{\rm diag.} V_u^{R\dagger}$, $M_d=V_d^{L} M_{d}^{\rm diag.} V_d^{R\dagger}$. In this new basis, where the Higgs and quark mass matrices are all diagonal, $\Gamma_u(m_Z),\ \Gamma_d(m_Z)$ are not diagonal and thus give rise to the following flavour violating neutral Higgs couplings:
\beq
\begin{split}
\mathcal{L}\supset\ &\bar{u}_L\Delta_u\left[\cos(\alpha-\beta)h + \sin(\alpha-\beta)H -iA\right]u_R\\
&+\bar{d}_L\Delta_d\left[\cos(\alpha-\beta)h + \sin(\alpha-\beta)H +iA\right]d_R,
\end{split}
\eeq
where:
\bea
\Delta_u&=&\frac{1}{v}V_u^{L\dagger}(m_Z)\Gamma_u(m_Z)V_u^R(m_Z),\label{Du}\\
\Delta_d&=&\frac{1}{v}V_d^{L\dagger}(m_Z)\Gamma_d(m_Z)V_d^R(m_Z).\label{Dd}
\eea
It is possible to calculate approximate expressions for $\Delta_u$,
$\Delta_d$ noting that:
\bea
\Delta_u &=&\frac{1}{v}(V_u^{L\dagger}\Gamma_u  M_u^{-1} V_u^L)
(V_u^{L\dagger}M_u V_u^R),
\eea
and analogously for $\Delta_d$. Substituting eqs.~\eqref{EWcoupling}, \eqref{M-at-MZ} and \eqref{Gamma-at-MZ} and keeping the lower order terms in $\epsilon_{u}$, $\delta_u$ we find that the off-diagonal couplings read:
\bea
\Delta_u^{\rm off-diag.}=E_uQ_u\label{Delta_u_off},\\
\Delta_d^{\rm off-diag.}=E_dQ_d\label{Delta_d_off},\label{DeltaEQ}
\eea
where, assuming real $\psi_u, \psi_d$:
\begin{align}
Q_u&\equiv \frac{1}{v^3}\left(V_{CKM}\left(M_{d}^{diag.}\right)^2V_{CKM}^\dagger M _{u}^{diag.}\right)^{\rm off-diag.},\\
E_u &\equiv \frac{1}{8 \pi^2} \frac{\sin (2(\psi_u -\psi_d))}{\cos^2(\beta-\psi_u)\cos^2(\beta-\psi_d)}\log\left( \frac{m_Z}{\Lambda}\right)\label{Eu},\\
Q_d&\equiv \frac{1}{v^3} \left(V_{CKM}^\dagger \left(M_{u}^{diag.}\right)^2V_{CKM}M_{d}^{diag.}\right)^{\rm off-diag.},\\
E_d&\equiv -E_u\label{Ed}.
\end{align}
Thus, the off-diagonal elements of the flavour violating Higgs couplings $\Delta_{u,d}$ can be factorized in two parts: $Q_{u,d}$ are determined by the experimental values of the entries of the CKM matrix and the quark masses, whereas $E_{u,d}$ depend on the unknown details of the 2HDM and the scale $\Lambda$ at which the alignment condition is imposed. It is apparent from \eqref{Eu} and \eqref{Ed} that $E_{u,d}$ depend just on two parameters: $\beta-\psi_u$ and $\beta-\psi_d$. Moreover, for $\psi_u=\psi_d$ and $\psi_u=\psi_d \pm \pi/2$ the flavour violating Higgs couplings vanish, since $E_u = E_d =0$. This choice includes as special cases the type I ($\psi_u=\psi_d=0$) and type II ($\psi_u=0,\ \psi_d=\pi/2$) 2HDMs.

\begin{figure}
\subfigure[Analytical approximation of $E_d$]{
\centering
\epsfig{figure=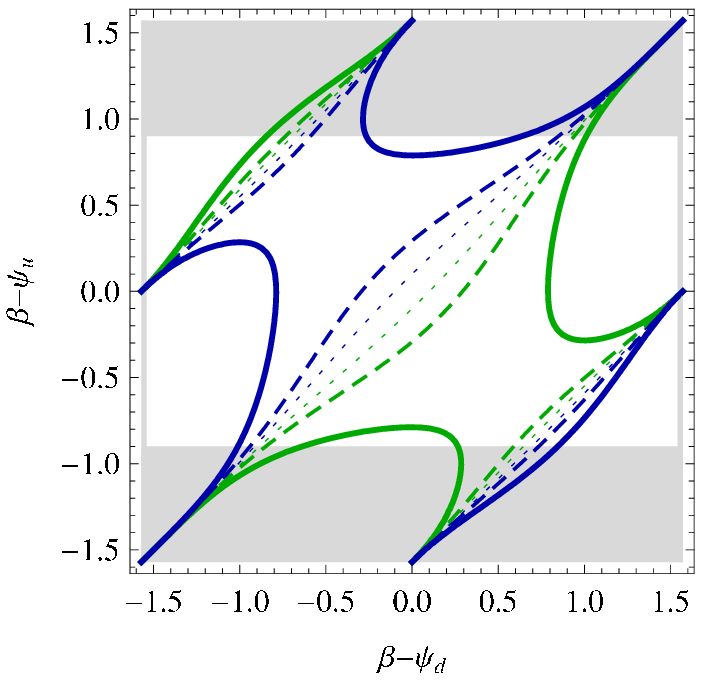,width=70mm}
\label{plotAnal}
}\hfill
\subfigure[Numerical result for $2.5 E_d$]{
\centering
\epsfig{figure=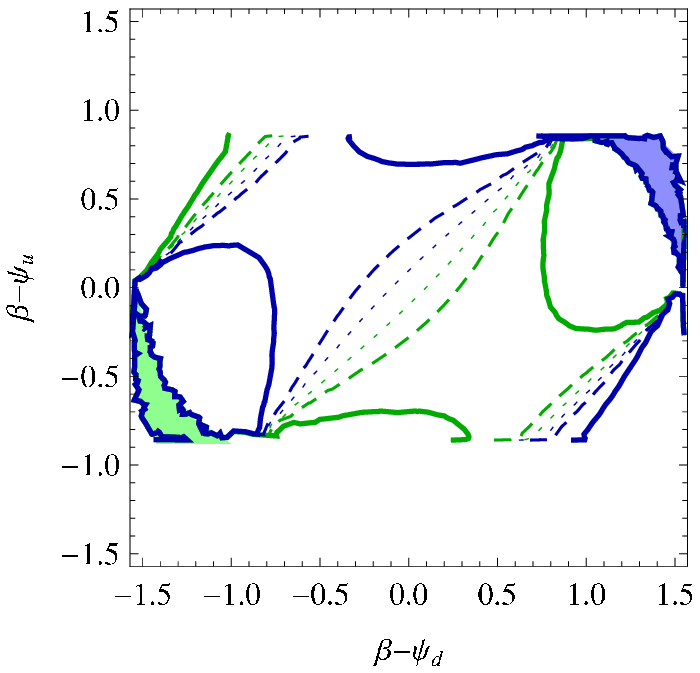,width=70mm}
\label{plotNum}
}
\caption{Contour plots of $E_d$ for $\Lambda=10^{19}{\rm ~GeV}$.
The left figure corresponds to the analytic formula, eq.~\eqref{Ed}.
Solid/dashed/dotted lines correspond to the absolute values of 1/0.3/0.1,
blue lines correspond to negative values, green lines to positive ones.
The right figure shows $2.5 \Delta_{d,23}/Q_{d,23}$ where $\Delta_{d,23}$ has been obtained by numerically solving the RGEs. The rescaling was done in order to make the comparison to the analytical result easier.}\label{plotEd}
\end{figure}

In fig.~\ref{plotEd} contours of $E_d$ are plotted.
The parameter range $-\frac{\pi}{2}<\beta-\psi_{d,u}<\frac{\pi}{2}$ is sufficient as $E_d$ is invariant under the shift $\beta -\psi_{d,u} \to \beta -\psi_{d,u} +\pi$.
As cut-off $\Lambda=10^{19}{\rm ~GeV}$ has been chosen as it is the scenario where maximal FCNCs can be expected and largest deviations of the leading log approximation. For this cut-off the grey shaded rectangles are not accessible as some Yukawa couplings become non-perturbative below $\Lambda=10^{19}{\rm ~GeV}$.
As will be shown in section~\ref{Experiments} $E_d$ can easily be $\mathcal{O}(1)$ while still satisfying the bounds on exotic contributions to the FCNCs.

To evaluate the accuracy of the analytical formulae we have calculated the flavour violating Higgs coupling $\Delta_{d\,23}$ solving numerically the full one-loop RGEs in the appendix.\footnote{We have used $\beta-\psi_e=0$ but the result is independent  unless $\cos(\beta-\psi_e) \to 0$.} We show in fig.~\ref{plotNum} $\Delta_{d\,23}/Q_{d\,23}$ and we find a very good agreement with the value $E_d$ calculated analytically in eqs.~\eqref{Eu} and \eqref{Ed} up to an overall factor of 2.5 (due to the large, flavour independent, effects in the running of the strong coupling constant and the top Yukawa coupling, which are not contemplated by the leading log approximation). We find, nevertheless, a new feature: There are regions with flipped sign at the top right and bottom left of the figure which are shown shaded.
In these regions, differences in the running of $Y_d^{(1)}$ compared to $Y_d^{(2)}$, not present in the leading log approximation, lead to a change of the sign of $M_d$, cf.\ eq.~\eqref{M-at-MZ}. Diagonalizing the quark masses, eqs.~\eqref{dVd}, this sign is transferred to $\Delta_d$.

Before deriving upper bounds on the flavour off-diagonal couplings we will present our approximate formulae in two parameterizations widely used in the literature: the Wolfenstein parameterization of the CKM matrix and the Cheng \& Sher parameterization of flavour violating couplings in a general 2HDM.

\paragraph{In the Wolfenstein parameterization}

Using the Wolfenstein parameterization of the CKM matrix \cite{Wolfenstein:1983yz}:
\beq
V_{CKM}=
\left(\begin{array}{ccc}
1-\frac{\lambda^2}{2} & \lambda & A \lambda^3 (\rho-i\eta)\\
-\lambda & 1-\frac{\lambda^2}{2} & A\lambda^2\\
A\lambda^3(1-\rho-i\eta) & -A\lambda^2 & 1
\end{array}\right)
+\mathcal{O}(\lambda^4)
\eeq
with $\lambda \approx 0.2$, $A\approx 0.8$, $\rho\approx 0.2$ and $\eta\approx 0.3$ \cite{Amsler:2008zzb} and the approximate expressions
\beq
M_{u}^{diag.}\sim\frac{v}{\sqrt{2}}{\rm diag}(\lambda^6,\lambda^3,1),\ \ \ M_{d}^{diag.}\sim\frac{v}{\sqrt{2}}{\rm diag}(\lambda^6,\lambda^4,\lambda^2),
\eeq
we get the following estimates:
\begin{align}
[Q_u]_{12}&\sim\lambda^{12}, & 
[Q_u]_{13}&\sim\lambda^8, &
[Q_u]_{23}&\sim\lambda^6,\label{wolfiu}\\
[Q_d]_{12}&\sim\lambda^9, &
[Q_d]_{13}&\sim\lambda^5,  &
[Q_d]_{23}&\sim\lambda^4,\label{wolfid}
\end{align}
and smaller values for the $(21), (31), (32)$ entries.

\paragraph{In the Cheng \& Sher parameterization}

The non-diagonal couplings of 2HDM are often parameterized as \cite{Cheng:1987rs}:
\begin{equation}
(\Delta_{u})_{ij}=\lambda^u_{ij}\frac{\sqrt{m_{u\,i} m_{u\,j}}}{v},~~~~~~~~~
(\Delta_{d})_{ij}=\lambda^d_{ij}\frac{\sqrt{m_{d\,i} m_{d\,j}}}{v}.
\label{ChengSher}
\end{equation}
Bounds on the coefficients $\lambda_{ij}^{u,d}$ have been derived from experimental results e.g.\ in \cite{Atwood:1996vj}. These bounds depend on the masses of the Higgs bosons and on whether the parameters $\lambda_{ij}^{u,d}$ are assumed to be universal or to posses some kind of hierarchy. 
We do \textit{not} assume universality.
Instead the Yukawa alignment condition leads to:
\begin{align}
|\lambda^u_{12}|&\sim 6\times 10^{-7}\ E_u, &
|\lambda^u_{13}|&\sim  10^{-4}\ E_u, & 
|\lambda^u_{23}|&\sim  7\times 10^{-5}\ E_u,\label{Chengu}\\
|\lambda^d_{12}|&\sim 5\times 10^{-4}\ E_d, & 
|\lambda^d_{13}|&\sim  6\times 10^{-2}\ E_d, &
|\lambda^d_{23}|&\sim  0.1\ E_d.\label{Chengd}
\end{align}

\section{Experimental Bounds}
\label{Experiments}

In general, there are numerous experimental bounds on the parameters of 2HDMs. Constraints derived for the type I or type II apply also in our scenario as it is more general. In \cite{Pich:2009sp} bounds on the aligned 2HDM have been studied explicitly. As the alignment condition is broken by radiative corrections, in addition tree-level FCNCs are present. This leads to further constraints on the parameters of this type of 2HDMs, as FCNCs are known from experiment to be highly suppressed.

\subsection{Meson-antimeson mixing}

\begin{figure}
\subfigure[$B_{s}-\bar{B}_{s}$ mixing]{
\begin{fmffile}{2HDMmixingTreeB}
\fmfframe(10,0)(0,5){
\begin{fmfgraph*}(50,20)
\fmfleft{i1,i2}
\fmfright{o1,o2}
\fmf{fermion}{i1,v1}
\fmflabel{$s$}{i1}
\fmf{scalar, label=$h,,H,,A$}{v1,v2}
\fmf{fermion}{v1,i2}
\fmflabel{$b$}{i2}
\fmf{fermion}{v2,o1}
\fmflabel{$b$}{o1}
\fmf{fermion}{o2,v2}
\fmflabel{$s$}{o2}
\end{fmfgraph*}}
\end{fmffile}
\label{2HDMBBbarTree}
}
\subfigure[$\bar{B}_{s} \to \mu^+ \mu^-$ ]{
\centering
\begin{fmffile}{2HDMrareBll}
\fmfframe(10,)(0,5){
\begin{fmfgraph*}(50,20)
\fmfleft{i1,o1}
\fmfright{i2,o2}
\fmf{fermion}{i1,v1}
\fmflabel{$s$}{i1}
\fmf{fermion}{v1,o1}
\fmflabel{$b$}{o1}
\fmf{scalar,label=$h,,H,,A$}{v1,v2}
\fmf{fermion}{i2,v2}
\fmflabel{$\mu^+$}{i2}
\fmf{fermion}{v2,o2}
\fmflabel{$\mu^-$}{o2}
\end{fmfgraph*}}
\end{fmffile}
\label{2HDMrareBTree}
}
\caption{Feynman diagrams for tree level flavour changing processes in a general 2HDM }
\end{figure}
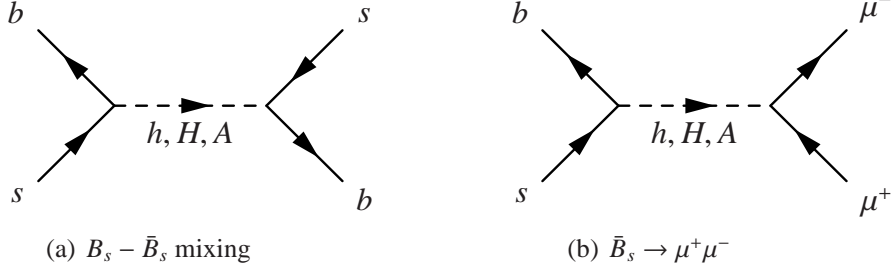

Stringent experimental bounds on FCNCs come from meson-antimeson mixing. In the SM this mixing can occur only at loop level while in a general 2HDM there is also a tree level mediation, see fig.~\ref{2HDMBBbarTree}.
As the flavour eigenstates are thus not mass eigenstates this mixing leads to tiny mass differences that have been determined experimentally for $B_d^0,B_s^0,D^0$ and $K^0$ mesons.
Here, we treat only the $B_{s}^0-\bar{B}_{s}^0$ system as it gives the strongest constraints (see e.g.\ eqs.~\eqref{Chengu} and \eqref{Chengd}). The effective Hamiltonian of the $\Delta B = 2$ transition $B_s^0\leftrightarrow\bar{B}^0_s$ is at scale $\sim m_Z$:
\beq
H_{\rm eff.}^{\Delta B=2}=\sum_{i,a}C_i^a(m_Z)Q_i^a(m_Z),
\eeq
where in a 2HDM with flavour violation at tree level the relevant operators are:
\bea
Q_1^{SLL}=(\bar{b}_R s_L)(\bar{b}_R s_L),&Q_1^{SRR}=(\bar{b}_L s_R)(\bar{b}_L s_R),&Q_2^{LR}=(\bar{b}_R s_L)(\bar{b}_L s_R).
\eea
The corresponding Wilson coefficients can be read off the effective Hamiltonian to be:
\begin{align}
C_1^{SLL}&=-\frac{(\Delta_{d\,23}^{*})^2}{2}\left(\frac{s^2_{\alpha\!-\!\beta}}{m_H^2}+\frac{c^2_{\alpha\!-\!\beta}}{m_h^2}-\frac{1}{m_A^2}\right) &
C_1^{SRR} &=-\frac{(\Delta_{d\,32})^2}{2}\left(\frac{s^2_{\alpha\!-\!\beta}}{m_H^2}+\frac{c^2_{\alpha\!-\!\beta}}{m_h^2}-\frac{1}{m_A^2}\right)\\
C_2^{LR}&=-\Delta_{d\,23}^*\Delta_{d\,32}\left(\frac{s^2_{\alpha\!-\!\beta}}{m_H^2}+\frac{c^2_{\alpha\!-\!\beta}}{m_h^2}+\frac{1}{m_A^2}\right)
\end{align}
and the meson-antimeson mass difference can be calculated as:
\beq
\Delta m_{B_s}=\left| \Delta m_{B_s}^{\rm SM}+\frac{2}{3}m_{B_s}F_{B_s}^2 \left[  P_2^{LR}C_2^{LR}(m_Z)+P_1^{SLL}\left(C_1^{SLL}(m_Z)+C_1^{SRR}(m_Z)\right) \right]\right|,
\eeq
where the coefficients $P_i^a$ include both the renormalization group evolution from the high scale $m_Z$ down to low energy $\sim m_{B_s}$ and the hadronization of the quarks to mesons. They can be calculated using the formulae in \cite{Buras:2001ra} and lattice QCD results from \cite{Laiho:2009eu}. For the $B^0_s-\bar{B}^0_s$-system we get $P_2^{LR}=3.0$ and $P_1^{SLL}=-1.9$. 
As $[\Delta_{d}]_{ij}\gg [\Delta_{d}]_{ji}$ for $j>i$ the term involving $C_1^{SRR}$ is always neglible, whereas $C_2^{LR}$ dominates only for degenerate Higgs masses or in the decoupling limit ($c_{\alpha\!-\!\beta}\rightarrow 0,\ m_H\approx m_A$). Therefore:
\bea
\Delta m_{B_s} &\simeq&\left| \Delta m_{B_s}^{\rm SM}+\frac{1}{3}m_{B_s}F_{B_s}^2 P_1^{SLL} \Delta_{d\,23}^{*2}\left(\frac{s^2_{\alpha\!-\!\beta}}{m_H^2}+\frac{c^2_{\alpha\!-\!\beta}}{m_h^2}-\frac{1}{m_A^2}\right)\right|,\\
\Delta m_{B_s} &\simeq &\left| \Delta m_{B_s}^{\rm SM}+\frac{2}{3}m_{B_s}F_{B_s}^2 P_2^{LR} \Delta_{d\,23}^*\Delta_{d\,32}\left(\frac{s^2_{\alpha\!-\!\beta}}{m_H^2}+\frac{c^2_{\alpha\!-\!\beta}}{m_h^2}+\frac{1}{m_A^2}\right)\right|\ \ \ \left( \substack{\text {\footnotesize \mbox{degenerate}}\\ \text {\footnotesize \mbox{masses or}}\\ \text {\footnotesize \mbox{decoupling limit}}} \right).
\eea
The SM prediction for the mass difference $\Delta m_{B_s}^{\rm SM}= (135\pm20)\cdot 10^{-13}$ GeV \cite{Lunghi:2007ak} agrees quite well with the experimental value $\Delta m_{B_s}^{\rm exp.}=(117.0\pm0.8)\cdot 10^{-13}$ GeV \cite{Amsler:2008zzb}. Nevertheless there is room for new physics as long as the new contribution to the mass difference is smaller than the theoretical uncertainty, i.e.\ $20\cdot 10^{-13}$ GeV.
Using eq.~\eqref{Delta_d_off} this leads to the approximate bound (we take $F_{B_s}=238.8\pm9.5$ MeV \cite{Laiho:2009eu}, $m_{B_s}=5.37$ GeV \cite{Amsler:2008zzb} and values for the quark masses at $m_Z$ from \cite{Fusaoka:1998vc}):
\begin{equation}
\left|\frac{s^2_{\alpha\!-\!\beta}}{m_H^2}+\frac{c^2_{\alpha\!-\!\beta}}{m_h^2}-\frac{1}{m_A^2}\right| |E_d|^2 \lesssim \frac{1}{(80\mathrm{~Gev})^2}.
\end{equation}
Thus, even for light Higgs masses, $\mathcal{O}(100\ \rm GeV)$, the present experimental constraints from meson-antimeson mixing allow $E_d$ to be of $\mathcal{O}(1)$. In the case of degenerate masses or the decoupling limit the bound is even weaker:
\begin{equation}
\left|\frac{s^2_{\alpha\!-\!\beta}}{m_H^2}+\frac{c^2_{\alpha\!-\!\beta}}{m_h^2}+\frac{1}{m_A^2}\right| |E_d|^2 \lesssim \frac{1}{(20\mathrm{~Gev})^2}.
\end{equation}

\subsection{Leptonic B decays}

The decay $\bar{B}_s\rightarrow l^+l^-\ (l=e,\mu,\tau)$, based on the flavour transition $b\rightarrow s$, is another example of a process that can be mediated at tree level in a general 2HDM (see fig.~\ref{2HDMrareBTree}) but neither in the SM nor in an aligned 2HDM. As the branching ratio depends on the lepton Yukawa coupling one expects the decay $\bar{B}_s\rightarrow \tau^+ \tau^-$ to be favoured. However, the produced $\tau$'s decay immediately to jets and leptons whose observed invariant mass will not reconstruct back to the mass of the B meson, so that these decays cannot be tagged in detectors \cite{Savage:1991qh}. In contrast, tagging $\bar{B}_s\rightarrow \mu^+ \mu^-$ is rather easy leading to an unparalleled bound on the branching ratio.
Integrating out the Higgs boson, the matrix element for the $h$ exchange is, assuming real $\Delta_{e\,22}$ (see Feynman rule \eqref{Feynman_h}):
\beq
\mathcal{M}_{\bar{B}_s\rightarrow\mu^+\mu^-}^h=\frac{1}{4}c_{\alpha-\beta}(\Delta_{d\,23}-\Delta_{d\,32}^*)\bra{\bar{B}_s}\bar{s}\gamma^5 b\ket{0}\frac{1}{m_h^2}\left(-\frac{m_\mu}{v}s_{\alpha-\beta}+c_{\alpha-\beta}\Delta_{e\,22}\right)\bar{\mu}\mu\label{hexchange},
\eeq
since
$\bra{\bar{B}_s}\bar{s}b\ket{0}$ is zero, as $\bar{B}_s$ is parity-odd, whereas $\bar{s}b$ is parity-even (see e.g.~\cite{Grossman:1996qj}).  On the other hand, the matrix element for $H$ exchange is of the same form as eq.~\eqref{hexchange}, with the replacements $c_{\alpha-\beta}\rightarrow s_{\alpha-\beta},\ s_{\alpha-\beta}\rightarrow -c_{\alpha-\beta},\ m_h\rightarrow m_H$. Lastly, the invariant amplitude for the exchange of a pseudo-scalar $A$ can be inferred from eq.~\eqref{Feynman_A} to be:
\beq
\mathcal{M}_{\bar{B}_s\rightarrow\mu^+\mu^-}^A=\frac{1}{4}(\Delta_{d\,23}+\Delta_{d\,32}^*)\bra{\bar{B}_s}\bar{s}\gamma^5 b\ket{0}\frac{1}{m_A^2}\Delta_{e\, 22}\ \bar{\mu}\gamma^5\mu.
\eeq
The decay rate can now be straightforwardly calculated from the decay amplitudes.
Using
\beq
\bra{\bar{B}_s} \bar{s}\gamma^5b\ket{0}\approx if_Bm_{B_s},
\eeq
and neglecting again terms proportional to $[\Delta_{d}]_{ji, j>i}$ we obtain:
\begin{multline}
\Gamma_{\bar B_s \rightarrow \mu^+ \mu^-} =\frac{f_{B_s}^2m_{B_s}^3}{64 \pi} |\Delta_{d\,23}|^2\ \times\\
\left\{
\frac{\Delta_{e\,22}^2}{m_A^4} +
\left|
\frac{s_{\alpha\!-\!\beta}}{m_H^2}\left(\frac{m_\mu}{v}c_{\alpha\!-\!\beta}+\Delta_{e\,22} s_{\alpha\!-\!\beta}\right)+
\frac{c_{\alpha\!-\!\beta}}{m_h^2}\left(-\frac{m_\mu}{v}s_{\alpha\!-\!\beta}+\Delta_{e\,22} c_{\alpha\!-\!\beta}\right)
\right|^2
\right\}.
\end{multline}
Requiring that the tree level alone does not exceed the present experimental bound ${\rm BR}(\bar{B}_s\rightarrow \mu^+ \mu^-) < 4.7\cdot 10^{-8}$ \cite{Amsler:2008zzb} we find for $\Delta_{e\,22} \gg \frac{m_\mu}{v}$:
\begin{equation}
\sqrt{
\frac{1}{m_A^4}  +
\left|
\frac{s_{\alpha\!-\!\beta}^2}{m_H^2}+
\frac{c_{\alpha\!-\!\beta}^2}{m_h^2}
\right|^2
} \Delta_{e\,22}
|E_d| \lesssim \frac{1}{(600\mathrm{~Gev})^2}.
\end{equation}
Hence for Higgs masses of $\mathcal{O}(100\ \rm GeV)$ and $E_d$ of $\mathcal{O}(1)$ there is only a conflict for $\tan(\psi_e-\beta) \sim \Delta_{e\,22}\frac{v}{m_\mu} \gtrsim 50$.
The opposite case, $\Delta_{e\,22} \ll \frac{m_\mu}{v}$, results in a bound on $E_d$ and the Higgs masses only. However, for generic Higgs masses the bound coming from $B_s-\bar{B}_s$ mixing is stronger.

\section{Conclusions and outlook}
\label{Conclusions}

We have studied in this letter the general two Higgs doublet model imposing that the Yukawa couplings for each fermion type are aligned at a high energy cut-off scale. This hypothesis guarantees the absence of new sources of flavour violation at tree level. We have shown, however, that quantum corrections spoil in general the alignment condition, inducing small flavour violating Higgs couplings at low energies. Without imposing {\it ad hoc} discrete symmetries and in the absence of tunings, the radiatively induced flavour violation in the Higgs sector yields the minimal size of the exotic contributions to the FCNCs in any 2HDM. We have shown that, for a wide range of parameters, this exotic contribution is well below the experimental constraints from meson-antimeson mixing and leptonic $B$-decays. 

In this letter we have concentrated on the constraints in the hadronic sector although the analysis can be extended to the leptonic sector. In the minimal extension of the Standard Model with one extra Higgs doublet, the Yukawa alignment in the leptonic sector is preserved after including the radiative corrections, due to the exactly preserved family lepton numbers. However, this scenario is incompatible with the observed neutrino oscillations. In a forthcoming publication \cite{Braeuninger} we will analyse the implications for the lepton flavour violating processes of the 2HDM with heavy right-handed neutrinos, necessary to generate neutrino masses through the see-saw mechanism.

\subsection*{Acknowledgements}
We are grateful to Andrzej J. Buras, Gino Isidori and Tillmann Heidsieck for their valuable advice
in the calculation of the experimental constraints on the 2HDM from B physics. 
This work was partially supported 
by the DFG cluster of excellence ``Origin and Structure of the Universe''
and by the Graduiertenkolleg ``Particle Physics at the Energy Frontier
of New Phenomena''.

\paragraph*{Note added:}

Simultaneously to the submission of this manuscript, the related
works \cite{Buras:2010mh,Jung:2010ik} appeared. Our conclusions agree in the aspects
where the analyses overlap.

\begin{appendices}
\singlespacing
\section{RGEs of the Yukawa couplings}
\label{RGEformulae}

The Renormalization Group Equation of the Yukawa couplings reads
\begin{equation}
16\pi^2 \mu \frac{\diff Y_f}{\diff\mu}=\beta_{Y_f},
\end{equation}
where

\begin{eqnarray}
\beta_{Y_u^{(k)}}&=&a_u Y_u^{(k)} +\sum_{l=1,2}\left[3\Tr\left(Y_u^{(k)}Y_u^{(l)\dagger}+Y_d^{(k)\dagger}Y_d^{(l)}\right)+\Tr\left(Y_e^{(k)\dagger}Y_e^{(l)}\right)\right]Y_u^{(l)}\label{betaup}\\
&&+\sum_{l=1,2}\left(-2Y_d^{(l)}Y_d^{(k)\dagger}Y_u^{(l)}+Y_u^{(k)}Y_u^{(l)\dagger}Y_u^{(l)}+\frac{1}{2}Y_d^{(l)}Y_d^{(l)\dagger}Y_u^{(k)}+\frac{1}{2}Y_u^{(l)}Y_u^{(l)\dagger}Y_u^{(k)}\right),\nonumber\\
\beta_{Y_d^{(k)}}&=&a_d Y_d^{(k)} +\sum_{l=1,2}\left[3\Tr\left(Y_u^{(k)\dagger}Y_u^{(l)}+Y_d^{(k)}Y_d^{(l)\dagger}\right)+\Tr\left(Y_e^{(k)}Y_e^{(l)\dagger}\right)\right]Y_d^{(l)}\label{betadown}\\
&&+\sum_{l=1,2}\left(-2Y_u^{(l)}Y_u^{(k)\dagger}Y_d^{(l)}+Y_d^{(k)}Y_d^{(l)\dagger}Y_d^{(l)}+\frac{1}{2}Y_u^{(l)}Y_u^{(l)\dagger}Y_d^{(k)}+\frac{1}{2}Y_d^{(l)}Y_d^{(l)\dagger}Y_d^{(k)}\right),\nonumber\\
\beta_{Y_e^{(k)}}&=&a_e Y_e^{(k)} +\sum_{l=1,2}\left[3\Tr\left(Y_u^{(k)\dagger}Y_u^{(l)}+Y_d^{(k)}Y_d^{(l)\dagger}\right)+\Tr\left(Y_e^{(k)}Y_e^{(l)\dagger}\right)\right]Y_e^{(l)}\label{betaelectron}\\
&&+\sum_{l=1,2}\left(Y_e^{(k)}Y_e^{(l)\dagger}Y_e^{(l)}+\frac{1}{2}Y_e^{(l)}Y_e^{(l)\dagger}Y_e^{(k)}\right),\nonumber
\end{eqnarray}
where $a_f$ ($f=u,d,e$) stands for contributions due to gauge interactions, which are flavour-diagonal:
\begin{align}
a_u&=-8 g_s^2-\frac{9}{4}g^2-\frac{17}{12}g'^2,\\
a_d&=-8 g_s^2-\frac{9}{4}g^2-\frac{5}{12}g'^2,\\
a_e&=-\frac{9}{4}g^2-\frac{15}{4}g'^2,
\end{align}
where $g_s,\ g$ and $g'$ are the gauge couplings constants of ${\rm SU}(3)_C,\ {\rm SU}(2)_L$ and ${\rm U}(1)_Y$, respectively.
The terms in the first sum in the $\beta$-function are due to the Higgs wave function renormalization, the first term in the second sum is due to the vertex renormalization (absent for leptons) and the last three (two) terms are due to the renormalization of the fermion wave function.

\section{Yukawa couplings at the EW scale}
\label{lowscaleYukawa}

\subsection{d-quarks}

\begin{equation}
Y_d^{(k)}(m_Z)\approx k_d^{(k)}Y_d+\epsilon_d^{(k)}Y_uY_u^{\dagger}Y_d+\delta_d^{(k)}Y_dY_d^{\dagger}Y_d
\end{equation}
with
\begin{align}
k_d^{(1)}&=\cos\psi_d+\frac{\log\frac{m_Z}{\Lambda}}{16\pi^2}
\Big[\begin{aligned}[t]
  &a_d \cos\psi_d+3\cos\psi_u^*\cos(\psi_u-\psi_d)\Tr(Y_u^\dagger Y_u)\\
  &+3\cos\psi_d\Tr(Y_dY_d^\dagger)+\cos\psi_e\cos(\psi_e^*-\psi_d)\Tr(Y_eY_e^\dagger)   \Big],
 \end{aligned}\\
\epsilon_d^{(1)}&=\frac{\log\frac{m_Z}{\Lambda}}{16\pi^2}\left(\frac{1}{2}\cos\psi_d-2\cos\psi_u^*\cos(\psi_u-\psi_d)\right),\\
\delta_d^{(1)}&=\frac{3 \log\frac{m_Z}{\Lambda}}{32\pi^2}\cos\psi_d,
\end{align}
\begin{align}
k_d^{(2)}&=\sin\psi_d+\frac{\log\frac{m_Z}{\Lambda}}{16\pi^2}
\Big[\begin{aligned}[t]
 &a_d \sin\psi_d+3\sin\psi_u^*\cos(\psi_u-\psi_d)\Tr(Y_u^\dagger Y_u)\\
 &+3\sin\psi_d\Tr(Y_dY_d^\dagger)+\sin\psi_e\cos(\psi_e^*-\psi_d)\Tr(Y_eY_e^\dagger)    \Big],
\end{aligned}\\
\epsilon_d^{(2)}&=\frac{\log\frac{m_Z}{\Lambda}}{16\pi^2}\left(\frac{1}{2}\sin\psi_d-2\sin\psi_u^*\cos(\psi_u-\psi_d)\right),\\
\delta_d^{(2)}&=\frac{3 \log\frac{m_Z}{\Lambda}}{32\pi^2}\sin\psi_d;
\end{align}

\subsection{u-quarks}

\begin{equation}
Y_u^{(k)}(m_Z)\approx k_u^{(k)}Y_u+\epsilon_u^{(k)}Y_dY_d^{\dagger}Y_u+\delta_u^{(k)}Y_uY_u^{\dagger}Y_u
\end{equation}
with
\begin{align}
k_u^{(1)}&=\cos\psi_u+\frac{\log\frac{m_Z}{\Lambda}}{16\pi^2}
\Big[\begin{aligned}[t]
 &a_u \cos\psi_u+3\cos\psi_u\Tr(Y_u^\dagger Y_u)+3\cos\psi_d^*\cos(\psi_d-\psi_u)\Tr(Y_dY_d^\dagger)\\
 &+\cos\psi_e^*\cos(\psi_e-\psi_u)\Tr(Y_eY_e^\dagger)    \Big],
\end{aligned}\\
\epsilon_u^{(1)}&=\frac{\log\frac{m_Z}{\Lambda}}{16\pi^2}\left(\frac{1}{2}\cos\psi_u-2\cos\psi_d^*\cos(\psi_d-\psi_u)\right),\\
\delta_u^{(1)}&=\frac{3\log\frac{m_Z}{\Lambda}}{32\pi^2}\cos\psi_u,
\end{align}
\begin{align}
k_u^{(2)}&=\sin\psi_u+\frac{\log\frac{m_Z}{\Lambda}}{16\pi^2}
\Big[\begin{aligned}[t]
 &a_u \sin\psi_u+3\sin\psi_u\Tr(Y_u^\dagger Y_u)+3\sin\psi_d^*\cos(\psi_d-\psi_u)\Tr(Y_dY_d^\dagger)\\
 &+\sin\psi_e^*\cos(\psi_e-\psi_u)\Tr(Y_eY_e^\dagger)   \Big],
\end{aligned}\displaybreak[2]\\
\epsilon_u^{(2)}&=\frac{\log\frac{m_Z}{\Lambda}}{16\pi^2}\left(\frac{1}{2}\sin\psi_u-2\sin\psi_d^*\cos(\psi_d-\psi_u)\right),\displaybreak[2]\\
\delta_u^{(2)}&=\frac{3\log\frac{m_Z}{\Lambda}}{32\pi^2}\sin\psi_u;
\end{align}

\subsection{Leptons}
As there are no vertex corrections in the leptonic sector, the coupling at the electroweak scale has a  different structure:
\begin{equation}
Y_e^{(k)}(m_Z)\approx k_e^{(k)}Y_e+\delta_e^{(k)}Y_eY_e^{\dagger}Y_e
\end{equation}
with
\begin{align}
k_e^{(1)}&=\cos\psi_e+\frac{\log\frac{m_Z}{\Lambda}}{16\pi^2}
\Big[\begin{aligned}[t]
 &a_e \cos\psi_e+3\cos\psi_u^*\cos(\psi_u-\psi_e)\Tr(Y_u^\dagger Y_u)\\
 &+3\cos\psi_d\cos(\psi_d^*-\psi_e)\Tr(Y_dY_d^\dagger)+\cos\psi_e\Tr(Y_eY_e^\dagger)\Big],
\end{aligned}\\
\delta_e^{(1)}&=\frac{3\log\frac{m_Z}{\Lambda}}{32\pi^2}\cos\psi_e,
\end{align}
\begin{align}
k_e^{(2)}&=\sin\psi_e+\frac{\log\frac{m_Z}{\Lambda}}{16\pi^2}
\Big[\begin{aligned}[t]
 &a_e \sin\psi_e+3\sin\psi_u^*\cos(\psi_u-\psi_e)\Tr(Y_u^\dagger Y_u)\\
 &+3\sin\psi_d\cos(\psi_d^*-\psi_e)\Tr(Y_dY_d^\dagger)+\sin\psi_e\Tr(Y_eY_e^\dagger)\Big],
\end{aligned}\\
\delta_e^{(2)}&=\frac{3\log\frac{m_Z}{\Lambda}}{32\pi^2}\sin\psi_e;
\end{align}

\section{Feynman rules}
\label{FeynmanRules}
\parindent0mm

\begin{minipage}{0.2\textwidth}
\begin{fmffile}{appendixHneu}
\fmfframe(3,3)(3,3){
\begin{fmfgraph*}(20,20)
\fmfleft{i1}
\fmfright{o1,o2}
\fmf{scalar, label=$H$}{i1,v1}
\fmf{fermion}{o1,v1}
\fmfv{label=$u_j,,d_j,,e_j$,label.angle=180}{o1}
\fmf{fermion}{v1,o2}
\fmfv{label=$u_i,,d_i,,e_i$,label.angle=180}{o2}
\end{fmfgraph*}}
\end{fmffile}
\end{minipage}
\begin{minipage}{0.8\textwidth}
\begin{flalign}\label{Feynman_H}
=-\frac{i}{2}\left[2c_{\alpha-\beta}\frac{m_{f\,i}}{v} \delta_{ij}
+s_{\alpha-\beta}\left((\Delta_f+\Delta_f^\dagger)_{ij}+(\Delta_f-\Delta_f^\dagger)_{ij}\gamma^5\right)\right] &&
\end{flalign}
for $f=u,d,e$
\end{minipage}

\vspace{0.5cm}

\begin{minipage}{0.2\textwidth}
\begin{fmffile}{appendixsmallh}
\fmfframe(3,3)(3,3){
\begin{fmfgraph*}(20,20)
\fmfleft{i1}
\fmfright{o1,o2}
\fmf{scalar, label=$h$}{i1,v1}
\fmf{fermion}{o1,v1}
\fmfv{label=$u_j,,d_j,,e_j$,label.angle=180}{o1}
\fmf{fermion}{v1,o2}
\fmfv{label=$u_i,,d_i,,e_i$,label.angle=180}{o2}
\end{fmfgraph*}}
\end{fmffile}
\end{minipage}
\begin{minipage}{0.8\textwidth}
\begin{flalign}\label{Feynman_h}
=-\frac{i}{2}\left[-2s_{\alpha-\beta}\frac{m_{f\,i}}{v} \delta_{ij}
+c_{\alpha-\beta}\left((\Delta_f+\Delta_f^\dagger)_{ij}+(\Delta_f-\Delta_f^\dagger)_{ij}\gamma^5\right)\right]&&
\end{flalign}
for $f=u,d,e$
\end{minipage}

\vspace{0.5cm}

\begin{minipage}{0.2\textwidth}
\begin{fmffile}{appendixA}
\fmfframe(3,3)(3,3){
\begin{fmfgraph*}(20,20)
\fmfleft{i1}
\fmfright{o1,o2}
\fmf{scalar, label=$A$}{i1,v1}
\fmf{fermion}{o1,v1}
\fmfv{label=$u_j,,d_j,,e_j$,label.angle=180}{o1}
\fmf{fermion}{v1,o2}
\fmfv{label=$u_i,,d_i,,e_i$,label.angle=180}{o2}
\end{fmfgraph*}}
\end{fmffile}
\end{minipage}
\begin{minipage}{0.8\textwidth}
\begin{flalign}\label{Feynman_A}
=\frac{1}{2}\left[
(\Delta_f-\Delta_f^\dagger)_{ij}+(\Delta_f+\Delta_f^\dagger)_{ij}\gamma^5\right]&&
\end{flalign}
for $f=d,e$ and for $f=u$ the same with negative sign
\end{minipage}

\vspace{0.5cm}

\begin{minipage}{0.2\textwidth}
\begin{fmffile}{appendixHp}
\fmfframe(3,3)(3,3){
\begin{fmfgraph*}(20,20)
\fmfleft{i1}
\fmfright{o1,o2}
\fmf{scalar, label=$H^+$}{i1,v1}
\fmf{fermion}{o1,v1}
\fmfv{label=$d_j,,e_j$,label.angle=180}{o1}
\fmf{fermion}{v1,o2}
\fmfv{label=$u_i,,\nu_i$,label.angle=180}{o2}
\end{fmfgraph*}}
\end{fmffile}
\end{minipage}
\begin{minipage}{0.8\textwidth}
\begin{flalign}
=-\frac{i}{\sqrt{2}}\left[
(-\Delta_u^\dagger V_{CKM}+V_{CKM}\Delta_d)_{ij}+(\Delta_u^\dagger V_{CKM}+V_{CKM}\Delta_d)_{ij}\gamma^5\right]&&
\end{flalign}
for quarks and $V_{CKM} \Delta_d \to \Delta_e$, $\Delta_u \to 0$ for leptons
\end{minipage}
\vspace{0.3cm}

Here $m_{f\,i}$ is the corresponding fermion mass and $\Delta_f$ is defined by eqs.~\eqref{Du}, \eqref{Dd} and analogously for the leptons.
\end{appendices}

\singlespacing

\bibliographystyle{utphys}
\bibliography{Article11}

\providecommand{\href}[2]{#2}\begingroup\raggedright\begin{thebibliography}{10}

\bibitem{D'Ambrosio:2002ex}
G.~D'Ambrosio, G.~F. Giudice, G.~Isidori, and A.~Strumia, ``{Minimal flavour
  violation: An effective field theory approach},''
  \href{http://dx.doi.org/10.1016/S0550-3213(02)00836-2}{{\em Nucl. Phys.} {\bf
  B645} (2002)  155--187},
\href{http://arxiv.org/abs/hep-ph/0207036}{{\tt arXiv:hep-ph/0207036}}.

\bibitem{Pich:2009sp}
A.~Pich and P.~Tuzon, ``{Yukawa Alignment in the Two-Higgs-Doublet Model},''
  \href{http://dx.doi.org/10.1103/PhysRevD.80.091702}{{\em Phys. Rev.} {\bf
  D80} (2009)  091702},
\href{http://arxiv.org/abs/0908.1554}{{\tt arXiv:0908.1554 [hep-ph]}}.

\bibitem{Ferreira:2010xe}
P.~M. Ferreira, L.~Lavoura, and J.~P. Silva, ``{Renormalization-group
  constraints on Yukawa alignment in multi-Higgs-doublet models},''
\href{http://arxiv.org/abs/1001.2561}{{\tt arXiv:1001.2561 [hep-ph]}}.

\bibitem{Wolfenstein:1983yz}
L.~Wolfenstein, ``{Parametrization of the Kobayashi-Maskawa Matrix},''
\href{http://dx.doi.org/10.1103/PhysRevLett.51.1945}{{\em Phys. Rev. Lett.}
  {\bf 51} (1983)  1945}.

\bibitem{Amsler:2008zzb}
{\bf Particle Data Group} Collaboration, C.~Amsler {\em et al.}, ``{Review of
  particle physics},''
\href{http://dx.doi.org/10.1016/j.physletb.2008.07.018}{{\em Phys. Lett.} {\bf
  B667} (2008)  1}.

\bibitem{Cheng:1987rs}
T.~P. Cheng and M.~Sher, ``{Mass Matrix Ansatz and Flavor Nonconservation in
  Models with Multiple Higgs Doublets},''
\href{http://dx.doi.org/10.1103/PhysRevD.35.3484}{{\em Phys. Rev.} {\bf D35}
  (1987)  3484}.

\bibitem{Atwood:1996vj}
D.~Atwood, L.~Reina, and A.~Soni, ``{Phenomenology of two Higgs doublet models
  with flavor changing neutral currents},''
  \href{http://dx.doi.org/10.1103/PhysRevD.55.3156}{{\em Phys. Rev.} {\bf D55}
  (1997)  3156--3176},
\href{http://arxiv.org/abs/hep-ph/9609279}{{\tt arXiv:hep-ph/9609279}}.

\bibitem{Buras:2001ra}
A.~J. Buras, S.~Jager, and J.~Urban, ``{Master formulae for Delta(F) = 2
  NLO-QCD factors in the standard model and beyond},''
  \href{http://dx.doi.org/10.1016/S0550-3213(01)00207-3}{{\em Nucl. Phys.} {\bf
  B605} (2001)  600--624},
\href{http://arxiv.org/abs/hep-ph/0102316}{{\tt arXiv:hep-ph/0102316}}.

\bibitem{Laiho:2009eu}
J.~Laiho, E.~Lunghi, and R.~S. Van~de Water, ``{Lattice QCD inputs to the CKM
  unitarity triangle analysis},''
  \href{http://dx.doi.org/10.1103/PhysRevD.81.034503}{{\em Phys. Rev.} {\bf
  D81} (2010)  034503},
\href{http://arxiv.org/abs/0910.2928}{{\tt arXiv:0910.2928 [hep-ph]}}.

\bibitem{Lunghi:2007ak}
E.~Lunghi and A.~Soni, ``{Footprints of the Beyond in flavor physics: Possible
  role of the Top Two Higgs Doublet Model},''
  \href{http://dx.doi.org/10.1088/1126-6708/2007/09/053}{{\em JHEP} {\bf 09}
  (2007)  053},
\href{http://arxiv.org/abs/0707.0212}{{\tt arXiv:0707.0212 [hep-ph]}}.

\bibitem{Fusaoka:1998vc}
H.~Fusaoka and Y.~Koide, ``{Updated estimate of running quark masses},''
  \href{http://dx.doi.org/10.1103/PhysRevD.57.3986}{{\em Phys. Rev.} {\bf D57}
  (1998)  3986--4001},
\href{http://arxiv.org/abs/hep-ph/9712201}{{\tt arXiv:hep-ph/9712201}}.

\bibitem{Savage:1991qh}
M.~J. Savage, ``{Constraining flavor changing neutral currents with $B \to
  \mu^{+} \mu^{-}$},''
\href{http://dx.doi.org/10.1016/0370-2693(91)90756-G}{{\em Phys. Lett.} {\bf
  B266} (1991)  135--141}.

\bibitem{Grossman:1996qj}
Y.~Grossman, Z.~Ligeti, and E.~Nardi, ``{$B \to\tau^+ \tau^- (X)$ decays: first
  constraints and phenomenological implications},''
  \href{http://dx.doi.org/10.1103/PhysRevD.55.2768}{{\em Phys. Rev.} {\bf D55}
  (1997)  2768--2773},
\href{http://arxiv.org/abs/hep-ph/9607473}{{\tt arXiv:hep-ph/9607473}}.

\bibitem{Braeuninger}
C.~B. Braeuninger, A.~Ibarra, and C.~Simonetto {\em in preparation}  .

\bibitem{Buras:2010mh}
A.~J. Buras, M.~V. Carlucci, S.~Gori, and G.~Isidori, ``{Higgs-mediated FCNCs:
  Natural Flavour Conservation vs. Minimal Flavour Violation},''
\href{http://arxiv.org/abs/1005.5310}{{\tt arXiv:1005.5310 [hep-ph]}}.

\bibitem{Jung:2010ik}
M.~Jung, A.~Pich, and P.~Tuzon, ``{Charged-Higgs phenomenology in the Aligned
  two-Higgs- doublet model},''
\href{http://arxiv.org/abs/1006.0470}{{\tt arXiv:1006.0470 [hep-ph]}}.

\end{thebibliography}\endgroup

\end{document}